\documentclass[pra,twocolumn]{revtex4}
\usepackage{graphics}

\draft
\begin{document}

\title{Dynamics of the Formation of Bright Solitary Waves of Bose-Einstein Condensates in Optical Lattices.}
\author{L. Plaja , J. San Rom\'an}
 \affiliation{Departamento de F\'\i sica Aplicada, Universidad de
Salamanca,\\ E-37008 Salamanca, Spain}
\date{\today}

\begin{abstract}
We present a detailed description of the formation of bright solitary waves in optical lattices. To this end, we have considered a ring lattice geometry with large radius. In this case, the ring shape does not have a relevant effect in the local dynamics of the condensate, while offering a realistic set up to implement experiments with conditions usually not available with linear lattices (in particular, to study collisions). Our numerical results suggest that the condensate radiation is the relevant dissipative process in the relaxation towards a self-trapped solution. We show that the source of dissipation can be attributed to the presence of higher order dispersion terms in the effective mass approach. In addition, we demonstrate that  the stability of the solitary solutions is linked with particular values of the  width of the wavepacket in the reciprocal space. Our study suggests that these critical widths for stability depend on the geometry of the energy band, but are independent of the condensate parameters (momentum, atom number, etc.). Finally, the non-solitonic nature of the solitary waves is evidenced showing their instability under collisions.
\end{abstract}

\maketitle

\section{Introduction}

One of the most relevant characteristics of Bose-Einstein condensates (BEC) is the coherent nature of their macroscopic wavepackets \cite{andre97A}.  This is particularly important in the case of BEC in optical lattices, where the continuous diffraction of the wave in the lattice distorts severely  the dynamics expected for a  free condensate. 

The behavior of BEC in optical lattices has drawn the attention of the research community almost since the first realizations of condensates in the laboratories \cite{ander98A}. Several experiments have demonstrated fundamental aspects of the lattice dynamics as Bloch oscillations \cite{morsc01A} as well as the progressive refinement of the techniques of loading and control of the condensate inside the lattice \cite{hecke02A}.  

The affinity between the non-linear Schr\"odinger equation (NSLE), which governs the mean field dynamics of the condensate, and the equation for the propagation of a light pulse in the slowly varying envelope approximation, suggests that several phenomena of non-linear optics can be reproduced in BEC. Among them, the formation of self-trapped (i.e. non dispersive) pulses \cite{haseg73A} is especially relevant, due to its applications in signal transmission. A fundamental property of the NLSE is its integrability \cite{zakha71A} and, as a consequence,  the existence of soliton solutions. Bright solitons, or more generally solitary waves, in BEC have been demonstrated experimentally for the case of attractive interactions \cite{khayk02A,strec02A} and dark solitons in the case of repulsive interactions \cite{burge99A,densc00A}. 

Bright solitons can be also obtained for repulsive interactions in condensates near the edge of the Brillouin zone of an optical lattice (gap solitons) \cite{carus02A,alfim02A,yulin03A} . The first experimental observation of this type of solitary waves has been reported very recently \cite{eierm04A}. Despite the interest of this matter, little theoretical effort has been addressed to understand the mechanism of spontaneous formation of the solitary wave from an initial wavepacket. Only very recently, some studies include matter radiation as a fundamental process accompanying solitary waves \cite{yulin03A}. It is the aim of this paper to clarify the essential points of this process with the aid of ab initio numerical solutions of the NLSE equation. Our calculations suggest that matter radiation is the fundamental relaxation process that transforms the initial wave to a solitary wavepacket. Accordingly to \cite{yulin03A}, the origin of the radiation is the disparity of the wavefunction with the exact solution of the Schr\"odinger equation. In this paper, we further clarify this point by identifying  a close link between the radiation process and the presence of high-order dispersive terms in the effective wave equation for the pulse envelope. These terms arise from the non-parabolic character of the energy band in the neighborhood of the packet's mean momentum. The  effect of higher order non-linear terms has been studied in the context of light propagating through optical fibers \cite{wai__90A} as a source of non-integrability of the NSLE. 

The numerical results will be interpreted with the help of an analysis based in the Wannier function picture. Wannier functions have been employed before connecting BEC dynamics in optical lattices and the Bose-Hubbard model \cite{jaksc98A} as well as for obtaining approximated models (see for instance \cite{alfim02B}) . However, it has been rarely used as a tool to understand the dynamical aspects of BEC in optical lattices as provided by the exact numerical integration of the NLSE. Our approach will be summarized in  section \ref{s:theory}, where the effective equation for the envelope of the wavepacket in the Wannier basis is derived, including the higher order dispersive terms usually neglected in soliton theory. We shall use this approach to understand the results of the integration of the NLSE in section \ref{s:results}. 

Our numerical computations are two-dimensional. For convenience, we shall use ring optical lattices instead of the usual linear ones. In addition to represent advantages for the numerical calculations, we believe that this geometry is also practical in experiments where the structure of the lattice changes with position, or when different momenta have to be imprinted to wave packets at different positions. Up to the authorsÕ knowledge, these ring optical lattices have not been yet created in the laboratories, although we find no fundamental difficulties in their construction (for instance in a similar way as proposed in this paper). We would like to emphasize that, since the radius of the ring is taken much larger than the wavepacket dimensions, the dynamical effects related to the shape of the waveguide are small, therefore it is reasonable to expect that the results described in this paper will also be applied in the case of other smooth geometries, as linear lattices (however the discussion of collisions would require periodic boundary conditions).

\section{Theory}
 \label{s:theory}

We consider a localized BEC released in an atomic waveguide at an initial time $t=0 ms$. Atomic waveguides of arbitrary shape have been created in laboratories by means of dipole-force confinement \cite{gorli01A,birk01A,bongs01A} or current carrying structures \cite{schmi00A,folma00A}.  In the case of condensates with low atomic density and strong transversal confinement, the dynamics of the wavepacket is essentially longitudinal, i.e. one-dimensional \cite{plaja02A}. In these circumstances, the transversal dynamics is frozen to the lowest energy  eigenstate. Naming $\left \{x_1,x_2,x_3 \right \}$, the coordinates along the transversal and axial directions of the waveguide respectively, the wavepacket factorizes as $ \Psi({\bf r}, t)= \xi_{12}(x_1, x_2) \xi_3(x_3,t)$. Where  $\xi_{12}(x_1, x_2) $ corresponds to the lowest energy eigenstate of the transversal Hamiltonian. Therefore, the axial dynamics can be described by a one-dimensional Gross-Pitaevskii equation, which results from the projection of the original 3D on the transverse state,
\begin{equation}
\label{eq:gp}
i \hbar {{\partial } \over {\partial t}} \xi_3(x_3,t) =  \left ( H_0+H_{nl}  \right ) \xi_3(x_3,t)
\end{equation}
where $H_0=- {{\hbar^2} \over {2 m}} \partial^2 / \partial x_3^2 + V(x_3)$, $H_{nl}=g_{1D} \left | \xi_3(x_3,t) \right |^2$, with $g_{1D} = (4 \pi \hbar^2 N a_s/m) \int \int dx_1 dx_2  \left|  \xi_{12}(x_1, x_2)  \right |^4$, $N$ being the number of atoms and $a_s$ the scattering length of the condensed element (in our case $^{87}Rb$). The potential $V(x_3)$ corresponds to the axial lattice. Equation (\ref{eq:gp}) is a one-dimensional equation along the coordinate $x_3$, and $\xi_3(x_3,t)$ is normalized to unity. For simplicity, therefore, in the following we shall drop the coordinate index.

In our study, the condensate is assumed to be released initially in a bare waveguide ($V(x)=0$, at $t=0 ms$),  with a particular momentum $\hbar k_0$. Afterwards,  the condensate expands freely during $0.5 ms$, before the lattice potential is switched on linearly in time. After $10 ms$, the optical lattice reaches its maximum value, and remains constant, $V(x)=V_0 sin^2(\pi x/a_0)$, for $t \ge 10.5 ms$ ($a_0$ being the intersite distance). Under these circumstances, we have found that the condensate is loaded into the lattice adiabatically enough, with most of the population (typically about $85 \%$) in the first band. In addition, the remaining population in the excited bands disperses rapidly over all momenta, reducing the amplitude probabilities accordingly. As a result, the relevant dynamics corresponds to the first band with little interaction with the excited population. In the following, we shall assume $\xi(x,t)$ to represent only the wavefunction in the lowest band, normalized accordingly with its corresponding fraction of the total population.

To understand the evolution of the condensate once the optical lattice has been generated, we find convenient to express the wavefunction either as a superposition of Wannier functions of the first band \cite{ziman}
\begin{equation}
\label{eq:wannier_exp}
\xi(x,t)=\sum_{\ell} e^{i k_0 \ell} f(\ell,t) a(x-\ell)
\end{equation}
with $\sum_{\ell}  | f(\ell,t) |^2=1$, or the alternative superposition of Bloch functions
\begin{equation}
\label{eq:bloch_exp}
\xi(x,t)= \int dk  e^{-i \epsilon(k+k_0) t/ \hbar } C_k(t) \psi_{k+k_0}(x)
\end{equation}
with $H_0 \psi_{k+k_0}(x)=\epsilon(k+k_0) \psi_{k+k_0}(x)$. These viewpoints are linked by the identity 
\begin{equation}
\label{eq:psi_exp}
\psi_{k+k_0}(x)={{1} \over {\sqrt{N}}} \sum_\ell e^{i(k+k_0) \ell} a_n(x-\ell)
\end{equation}
Using (\ref{eq:wannier_exp}) and  (\ref{eq:bloch_exp}), together with  the orthogonality condition $\int dx a^*(x-\ell) a(x-\ell') = \delta_{\ell,\ell'}$, we find
\begin{equation}
\label{eq:f_exp}
f(\ell,t)={{1} \over {\sqrt{N}}} \int dk e^{-i \epsilon(k+k_0) t/ \hbar } e^{i k \ell }C_k(t)
\end{equation}
Introducing Eqs. (\ref{eq:wannier_exp}) and (\ref{eq:bloch_exp}) into (\ref{eq:gp}), and neglecting the overlap between Wannier functions of adjacent sites (assumption of tight binding) we find
\begin{eqnarray}
\label{eq:sum_ell}
 & \sum_{\ell} & e^{i k_0 \ell} \left ( i \hbar {{\partial}\over{\partial t}} f(\ell,t) \right )a(x-\ell) = \nonumber \\
 & & \int dk  e^{-i \epsilon(k+k_0) t/ \hbar } \epsilon(k+k_0) C_k(t) \psi_{k+k_0}(x) + \nonumber \\
 & g_{1D} & \sum_{\ell} e^{ik_0\ell} \left | a(x-\ell) \right |^2  a(x-\ell)  \left | f(\ell ,t) \right |^2  f(\ell,t)
 \end{eqnarray}
In the cases discussed in this paper, the condensate wavepacket extends over some tens of lattice sites. Accordingly, the corresponding wavefunction in the reciprocal space is a narrow peak centered on $k_0$, so we can approximate
\begin{equation}
\label{eq:energia_exp}
 \epsilon(k+k_0) \simeq \epsilon(k_0) + \hbar v_g k + {{\hbar^2 k^2} \over {2 m^*}}- \alpha_3 k^3
 \end{equation}
 where $v_g= (1/\hbar) (\partial/\partial k) \epsilon |_{k_0}$ and $1/m^*=(1/\hbar^2) (\partial^2/\partial^2 k) \epsilon |_{k_0}$ correspond to the standard definitions of the group velocity and effective mass respectively.  Additionally we keep the higher order dispersion term $\alpha_3=-(1/3!)  (\partial^3/\partial^3 k) \epsilon |_{k_0}$, which would be neglected if dealing with condensates extended over a greater number of lattice sites. 
 
Substituting Eq. (\ref{eq:energia_exp}), and using (\ref{eq:psi_exp}) and (\ref{eq:f_exp}), the first term in the {\em rhs} of  (\ref{eq:sum_ell}) can be written as
\begin{eqnarray}
\int & dk  & e^{-i \epsilon(k+k_0) t/ \hbar } \epsilon(k+k_0) C_k(t) \psi_{k+k_0}(x) \simeq \nonumber \\
&  &  \sum_\ell a(x-\ell) e^{i k_0 \ell} \left \{ \epsilon(k_0) +  v_g (-i \hbar) {{\partial} \over {\partial \ell}} - {{\hbar^2} \over {2 m^*}} {{\partial^2} \over {\partial \ell^2}} \right . \nonumber \\
 & & -i  \left . \alpha_3 {{\partial^3} \over {\partial \ell^3}} \right \} f(\ell,t)
\end{eqnarray}
Using this form and the orthogonality of Wannier functions, we can project Eq. (\ref{eq:sum_ell}) on the Wannier state $e^{k_0 \ell} a(x-\ell)$ to give
\begin{eqnarray}
\label{eq:gp_eff}
 & &i  \hbar {{\partial}\over{\partial t}} f(\ell,t) = \nonumber \\
  & & \left \{ \epsilon(k_0) +  v_g (-i \hbar) {{\partial} \over {\partial \ell}} - {{\hbar^2} \over {2 m^*}} {{\partial^2} \over {\partial \ell^2}}  -i   \alpha_3 {{\partial^3} \over {\partial \ell^3}}  \right \} f(\ell,t) \nonumber \\
 & &+  \tilde{g}  \left | f(\ell ,t) \right |^2  f(\ell,t)
 \end{eqnarray}
 where $ \tilde{g} =g_{1D} \int dx |a(x-\ell)|^4 $. Defining adimensional space and time variables as $\eta=( \ell - v_g t)/\ell_0$ and $\tau=t/\tau_0$, with $\ell_0=\sqrt{\hbar \tau_0/ |m^*|}$ and $\tau_0$ a characteristic time, Eq. (\ref{eq:gp_eff}) can be rewritten as
 \begin{eqnarray}
\label{eq:gp_eff_adim}
i  {{\partial}\over{\partial \tau}}q(\eta,\tau) & =  &\left \{ E_0 \mp {{1} \over {2}} {{\partial^2} \over {\partial \eta^2}} - i  \beta {{\partial^3} \over {\partial \eta^3}}  \right \} q(\eta,\tau)+ \nonumber \\
  & & \left | q(\eta,\tau) \right |^2  q(\eta,\tau)
 \end{eqnarray}
 with $q(\eta,\tau)=\sqrt{\tilde{g} \tau_0/\hbar} f(\eta,\tau)$, $\beta=\alpha_3 |m^*|^{3/2}/\hbar^{5/2} \tau_0^{1/2}$, and $E_0=\epsilon(k_0) \tau_0/\hbar$. This equation contains a first correction to the dispersion resulting from the non-parabolic shape of the band structure around $k_0$. Since contains a third derivative it is expected to become important the wider the wavepacket is in the reciprocal space or, alternatively, the closer is $k_0$ to the zero-dispersion point. In fact, the propagation of light in monomode fibers follows an equation, which is formally the complex conjugate of (\ref{eq:gp_eff_adim}) \cite{wai__90A}. 
 
\section{Results and Discussion}
\label{s:results}

To test the above viewpoint and, particularly, the relevance of the higher dispersion terms in Eq. (\ref{eq:gp_eff_adim}), we have computed ab initio numerical solutions of the Gross-Pitaevskii equation over a range of parameters. Systematic calculations of the dynamics of propagating solitary waves in more than one dimension can be effectively speeded up with the choice of a waveguide geometry with periodic boundary conditions. On the other hand, our interest comprises also the study of solitary wave collisions. This kind of experiments require the formation of two solitary waves of different momenta, and represents a non-trivial  extension of the present experimental setups used in the formation of bright gap solitons, which consider the translation of the whole lattice instead of imprinting an initial momentum to the wavepacket.  To fulfill these requirements, without departing from a realistic situation,  we find advantages in proposing the use of ring-shaped optical lattices (see Fig. \ref{fig:ring}). These kinds of lattices are composed by a ring-shaped atomic waveguide of the same sort as those used in \cite{gorli01A,birk01A,bongs01A,schmi00A,folma00A}. The optical lattice itself may be generated by a periodic angular modulation in the transverse distribution of intensity of a laser (of incidence perpendicular to the lattice plane) using the appropriate mask.  As in \cite{bongs01A} the waveguide can be loaded in different locations using a properly shaped initial trap. Finally, different initial momenta can be communicated to wavepackets at every position by, for instance, a phase imprinting method  \cite{dobre99A, densc00A}. 

The ring-shaped atomic waveguide is defined by a potential
\begin{equation}
V_{wg}(r,\phi,z)= {{1} \over {2}} m \omega_z^2 z^2 + {{1} \over {2}} m \omega_r^2 (r-r_0)^2
\end{equation}
where in our calculations $\omega_z=2 \pi \times 700 Hz$, $\omega_r=2 \pi \times 50 Hz$ and $r_0=40 \mu m$.    Since $\omega_z>>\omega_r$, for low-density condensates we can consider the dynamics along the $z$ direction as frozen, and write in the same fashion as in section \ref{s:theory},  $\Psi({\bf r},t)=\xi_z(x) \xi(r, \phi,t)$,  $\xi_z(x)$ being the ground state of the harmonic oscillator in $z$ direction. Following  the steps in section \ref{s:theory}, we find the two dimensional counterpart of Eq. (\ref{eq:gp})
\begin{eqnarray}
\label{eq:gp2}
 i \hbar {{\partial} \over {\partial t}} \xi(r, \phi,t) =  -  {{\hbar^2} \over {2m}} \left( {{1} \over {r}} {{\partial} \over {\partial r}} + {{\partial^2} \over {\partial r^2}} + {{1} \over {r^2}} {{\partial^2} \over {\partial \phi^2}} \right)  \xi(r, \phi,t) + \nonumber \\
 \left ( {{1} \over {2}} m \omega_r^2 (r-r_0)^2 + V(z) + g_{2D} | \xi(r, \phi,t)|^2  \right )  \xi(r, \phi,t) \nonumber \\
 \left . \right.
 \end{eqnarray}
where $g_{2D} = (4 \pi \hbar^2 N a_s/m) \int dz \left|  \xi_z(z)  \right |^4$, and $ V(z)= V_0 \sin^2(\pi r_0 \phi/a_0) $ is the lattice potential. We take as initial time the moment in which the transference of the condensate to the waveguide has been accomplished. At $t=0 ms$, an angular momentum $L= \hbar 58$ is imprinted to the condensate. This corresponds imprinting a linear momenta $ \hbar k_0= L/r_0$ along the waveguide circle. Note that, although high momentum rotating condensates are unstable in three dimensions and decay into smaller vortices, the waveguide reduces the effective dimensionality of the problem to one dimension.  At $t=0.5 ms$ the lattice potential is grown linearly in time during $10 ms$. In all the cases shown in this paper, the potential has $128$ wells and, therefore, the lattice constant is $a_0=2 \pi r_0/128=1.96 \mu m$. From $t=10.5 ms$ on, the lattice is stationary and the wavepacket evolves almost one dimensionally inside the waveguide, in the presence of the optical lattice

All our numerical computations correspond to solutions of Eq. (\ref{eq:gp2}) and, therefore, are two-dimensional.  However, since for our choice of $\omega_r$ the radial size of the wavefunction in the waveguide is much smaller than the radius $r_0$, the assumption of frozen radial dynamics made in section \ref{s:theory} is reasonable. This gives us the possibility of interpreting the results of the numerical integration of Eq. (\ref{eq:gp2}) with the theory developed in section \ref{s:theory}, with $x_1=z$, $x_2=r$ and $x_3=r \phi \simeq r_0 \phi$.

We have performed a set of numerical calculations involving a range of different lattice potential strengths  $V_0$ whose associated frequencies range from $64 Hz$ to $256 Hz$,  number of atoms ranging from $50$ to $200$, and initial momenta $\hbar k_0=\hbar 58/r_0$ and $\hbar k_0=\hbar 64/r_0$. To our knowledge, the experiments with condensates with these low number of atoms are difficult (for instance repeated measurements will destroy them ). However, the solitons observed recently \cite{eierm04A} are composed by 300 atoms. The scalability of the problem to higher number of atoms could be achieved provided the lattice intersite distance is preserved (therefore the band structure) and the initial condensate size is larger in $x_3$ so that the atomic density at every site is kept similar (so that the tunneling rate is not changed).  However, phase fluctuations may play an important role in these scaled systems and, therefore, could change drammatically the whole scenery and prevent the formation of the solitary wave.

A typical process of the formation of bright BEC solitary wave is shown in Fig. \ref{fig:solit_formation} for a condensate of $100$ atoms. Figure \ref{fig:solit_formation}(a)  shows the evolution of the envelope of  the wavepacket projected into the Wannier functions of the first band ($f(\ell,t)$ in Eq. (\ref{eq:f_exp})). Since the initial momentum is close to the edge of the Brillouin zone, at $t=0 ms$ ($V_0=0$) approximately $20 \%$  of the wavepacket extends into the second zone. As a result, even in the case of a perfect adiabatic switch on of the lattice, part of the population is loaded in the excited band. Our numerical results suggest that the population in the excited band expands rapidly in space, its probability amplitude decreasing accordingly, and therefore affects little to the evolution of the population in the lowest band, where the solitary wave is formed. The relevant dynamics, therefore concerns the population of the first band. As an example of this, Figure \ref{fig:solit_formation}(b) shows the total atomic density $|\xi|^2$ at $t=180 ms$, the solitary wave corresponds to the pulse of part (a), while the background noise corresponds to the radiated population in the first band and to the population of the excited bands. After the lattice is switched on, the condensate wavepacket in the first band is found to evolve freely with a velocity corresponding to the group velocity $ \simeq v_g(k_0)$. The stabilization of the pulse shape becomes apparent at times greater than $t=120 ms$.  It is important to note the presence of a small trapped population near $r_0 \phi=250 \mu m$ in addition to the moving solitary wave.  This population is emitted by the wavepacket during the time before adopting the solitary shape. Our calculations point out that this radiated population remains trapped in the lattice wells, since its low density prevents Josephson tunneling between sites. 
This fact is supported by the inspection of Fig. \ref{fig:zb_solit_form} (a), where the time evolution of the population in the first band, depicted already in  Fig. \ref{fig:solit_formation}, is plotted now projected on the Bloch basis.  Also by the inspection of  Fig. \ref{fig:zb_solit_form} (b), where  the same projection is plotted, but now restricted to the narrowest spatial box enclosing the solitary wave at every time. In both plots, the solitary wave corresponds to the peak near the edge of the Brillouin zone, 
and the smaller peak at zero momentum corresponds to the trapped radiation. In part (b), this later peak appears only at $t=120 ms$, i.e. when the radiation takes place, and disappears at later times, since the solitary wave moves out from the region where the radiation is trapped. The differences between the peaks belonging to the solitary wave in (a) and (b) suggest that there is an additional radiated component, which follows the solitary pulse (i.e. they have similar momenta), but located at a different spatial region. A close inspection of figure \ref{fig:solit_formation} demonstrates that it can be attributed to small amplitude fluctuations at the edge of the solitary wave. In the case of condensates of higher density, as in \cite{yulin03A}, radiation trapping is not possible since its increased density allows for Josephson tunneling. In these cases, the radiation peak is found at momenta different from $0$.

Figure \ref{fig:zb_solit_form}(a) and (b) also show that the solitary pulse is not located exactly at the edge of the Brillouin zone, but at some particular point inside. This is a remarkable point in the light of the discussion of the previous section, since implies the presence of non-vanishing third-order derivatives in the equation of motion (\ref{eq:gp_eff_adim}). Note also, that this also the case of any solitary wave with non-zero group velocity. It is known that the presence of a third derivative in the non-linear Schr\"odinger equation prevents its integrability and, as a consequence, the appearance of solitons. In contrast,  it has been reported the presence of solitary waves of light in optical fibers, which radiate constantly with a rate decreasing in time \cite{wai__90A}. These sorts of solutions have a spectral distribution of the form shown in Fig. \ref{fig:zb_solit_form}.

From our calculations, therefore, it becomes apparent that radiation plays a fundamental role in the process of formation of solitary waves, since it represents a dissipative process that forces the convergence to a stable self-trapped solution. We find important to stress that, in contrast to other alternative schemes for bright soliton formation (for instance changing the sign of the non linear parameter using Feshbach resonances), the presence of higher order dispersion is particular to the case of optical lattices, and provides a natural dissipative process leading to the convergence of the initial wavepacket into a solitary wave. 

The radiative process has also consequences in the dynamics of the center of mass of the condensate wave packet, as a result of the conservation of the overall momentum.  This phenomenon is reflected in our calculations, as it is shown in Fig. \ref{fig:k0_t}, where the time evolution of the momentum of the forming solitary wave is depicted as a function of time for two different initial momenta, $\hbar k_0=\hbar 58/r_0$ and $\hbar k_0=\hbar 64/r_0$. It should be stressed that the momenta are calculated from the mean value of the differential operator applied to the envelopes $f(r_0 \phi, t)$ of the lowest band. Since at $t=0 ms$ part of the wavepacket extends beyond the limits of the first Brillouin zone, the computed averaged momenta at this time is always smaller than $k_0$. The first case ($k_0=58/r_0$) corresponds to the same calculation as Fig. \ref{fig:solit_formation}, the formation of the solitary wave is described quite dramatically as a rapid stabilization of the wavepacket momentum.  An examination of the time evolution of the wavepacket, shows that the presence of a strong oscillating region before the solitary wave formation corresponds to the interaction of the wavepacket with its radiation, as one runs over the other. In addition, Fig. \ref{fig:k0_t} shows for this case an increase of the wavepacket momentum as the solitary wave is formed. As pointed out above, the low density of the radiated wave reduces drastically the Josephson tunneling rate. As a consequence, the momentum of the radiated wave is small, if not zero. In this situation, the momentum of the forming solitary wave should increase to account for momentum conservation, i.e. it will be accelerated. The case is rather different for the initial momentum $\hbar k_0=\hbar 64/r_0$, where the wavepacket momentum decreases rather than increases. We have found this to occur in situations in which  the initial wavepacket is closer to the edge of the Brillouin zone. In this circumstance, the matter is radiated mostly to the excited band, and not to the lowest band. This effect can be attributed to Landau-Zener tunneling induced by the nonlinear potential, which is more probable the smaller is the energy gap (i.e. at the edge of the Brillouin zone).    

As a general trend, we have found that the width of the condensate wavepacket in the reciprocal space is a critical parameter for stability.  Figure \ref{fig:k_width} shows the time evolution of this quantity for condensates with different initial number of atoms and for different initial momenta. In all these cases the formation of the solitary wave has its signature in the stabilization of the width of the wavepacket to a rather similar value. The same behavior is also present for other lattice potentials, however the convergence resulting to a different value.  This result seems to be a particular feature of the reciprocal width of the wavepacket, not shared by other parameters as the area of the soliton or its energy. We may find a physical explanation of this effect in equations (\ref{eq:energia_exp}) and  (\ref{eq:gp_eff}): from the previous discussion is natural to assume that the intensity of radiation depends on the relative importance of the higher order dispersion terms (third derivative) to the internal kinetic energy of the wavepacket (second derivative). Since the ratio between the third order and second order dispersion terms is $|2 m^* \alpha_3 \Delta k/ \hbar^2 |$, a narrower wavefunction in the reciprocal space will radiate at a slower rate than a wider one. The limiting width, therefore, can be considered as a situation in which radiation takes place at a asymptotic slow rate. We have found evidence of this in our calculations, which show that the final wavefunction has a minimum value of the dispersion ratio. Figure \ref{fig:ratio} shows these ratios as a function of time, for the same parameters as in Fig. \ref{fig:solit_formation}. The ratios have been calculated from the band structure according to the value of the effective mass and $\alpha_3$ at the instantaneous average momentum (shown in Fig  \ref{fig:k0_t}) and according to the instantaneous value of the reciprocal wavepacket width (Fig. \ref{fig:k_width}).

As pointed out above, one of the consequences of the presence of higher order dispersion terms is to prevent the integrability of the system. As a result, the self-trapped solutions do not correspond to solitons but to the less restrictive class of solitary waves. One of the strict test of the soliton nature is found in the elasticity of their collisions. As a consequence, solitons are known to preserve their shape under such events. In order to explore this dynamical aspect, we have taken advantage of the ring scheme to model a collision experiment. In such geometry, the transference of distinct  momenta to wavepackets located at different regions of the waveguide seems feasible, simply by tailoring the proper phase mask for the phase imprinting. Also, the periodic boundary allows the wavepacket to decay into a solitary wave in a small spatial region, which is helpful when computing numerically problems in dimensions higher than one. In Fig. \ref{fig:colision} we show the time evolution of two colliding solitary waves. The initial condition of the computation is a condensate splitted into two wavepackets, at symmetric sites in the ring, and imprinted with opposite momenta. Each of the wavepackets resembles in number of atoms and initial momentum with the case shown in Fig. \ref{fig:solit_formation}. Following the previous results of this paper, both wavepackets will decay into solitary waves at times close to $120 ms$. Therefore, the first plot in Fig. \ref{fig:colision} corresponds already to the two solitary waves in the ring heading one into each other.  As it is apparent from the figure, the collision is highly inelastic, leading to the disintegration of the solitary wavepackets.

\section{Conclusion}

We have analyzed the process of formation of solitary wavepackets of Bose-Einstein condensates loaded into optical lattices. Our results support the picture of  matter radiation acting as a dissipative process that leads to the decay of the initial wavefunction into the solitary shape. We demonstrate that the source of radiation is linked to the presence of higher order dispersion terms inherent to the lattice band structure. Our point of view is consistent with previous studies on the effect of these dispersive terms in the propagation of light in optical fibers. After the process has been identified, we have demonstrated that the center of mass dynamics of the wavepacket is closely connected with the form of radiation through the conservation of the overall momentum, resulting in the acceleration or deceleration of the wavepacket as the solitary wave is formed. By radiating part of its mass, the wavepacket shrinks in reciprocal space, reducing accordingly  the role of higher order dispersion terms. Consequently, radiation diminishes in time as the wavepacket evolves to a solitary wave of a fixed width in reciprocal space. This final width seems to be linked to the depth of the lattice sites and independent on the wavepacket parameters (initial momentum and number of atoms). Finally, we have studied the dynamics of collisions between these waves and found it highly inelastic, leading to the rapid dispersion of the solitary waves. All these results are supported by two-dimensional numerical integrations of the Gross-Pitaevskii equation in a ring-shaped optical lattice, and interpreted with the aid of a Wannier wavefunction analysis. The size of the ring is chosen large enough to avoid particular effects related with this geometry.

\pagebreak

\acknowledgments

We thank L. Santos for fruitful discussions. This work has been partially supported by the Spanish Ministerio de Ciencia y Tecnolog\' \i a (FEDER funds, grant BFM2002-00033) and by the Junta de Castilla y Le\'on (grant SA107/03).

\newpage

\begin{figure}
\caption{
Scheme for constructing ring-shaped optical lattices as proposed in the text: a laser (black arrow) is aimed perpendicularly to a ring-shaped atomic waveguide, after being transmitted through an angular modulated transmission plate. In our calculations, the ring radius is $40 \mu m$, enclosing 128 lattice sites.}
\label{fig:ring}
\end{figure}

\begin{figure}
\caption{(a) Plot of $| f(\ell, t) |^2$ (wavefunction's envelope in the Wannier basis of the lowest energy band) as a function of time and the position in the ring, showing how the initial wavepacket converges into a solitary wave. The parts included in the dashed box have been magnified to enhance visibility, and the arrows show the radiation component traveling with the pulse. (b) Plot of the atom density at $t=180 ms$. The condensate of $100$ atoms is initially released with momentum $\hbar 58/r_0$ along the ring circle, and the optical lattice is linearly built between $t=0.5 ms$ and $t=10.5 ms$, afterwards remains constant with $V_0/\hbar \simeq 128 Hz$. }
\label{fig:solit_formation}
\end{figure}

\begin{figure}
\caption{(a) Plot of the evolution of the atom population in the first band of the Brillouin zone (projection into Bloch states as given by Eq. (\ref{eq:bloch_exp})), for the same case as in Fig. \ref{fig:solit_formation}. (b) The same but using only the part of the wavefunction $\xi$ in the spatial box including only the region occupied by the solitary peak at every instant of time. Regions enclosed in the dashed box have been magnified. The reciprocal space momenta is given in units of $g=2 \pi/a_0=128/r_0$.}
\label{fig:zb_solit_form}
\end{figure}

\begin{figure}
\caption{Time evolution of the mean value of the reciprocal lattice momentum for the same parameters as in  Fig. \ref{fig:solit_formation} but initial momenta $\hbar k_0= \hbar 58/r_0$ (solid line) and $\hbar k_0=\hbar 64/r_0$ (i.e. at the edge of the Brillouin zone, dashed line).}
\label{fig:k0_t}
\end{figure}

\begin{figure}
\caption{Time evolution of the width of the wavefunction in the reciprocal space for different lattice depths: (a) $V_0/\hbar \simeq 128 Hz$, (b) $V_0/\hbar \simeq 256 Hz$, (c)) $V_0/\hbar \simeq 64 Hz$, and for condensates with different initial conditions (number of atoms and momentum), written as $\{N, k/g \}$  (a.1) $\{200, 0.453 \}$, (a.2) $\{100, 0.453 \}$, (a.3) $\{50, 0.453 \}$, (a.4) $\{100, 0.5 \}$, (b.1) $\{100, 0.453 \}$, (b.2) $\{50, 0.453 \}$, (c.1) $\{100, 0.453 \}$, (c.2) $\{50, 0.453 \}$.}
\label{fig:k_width}
\end{figure}

\begin{figure}
\caption{Ratio between the third and second order dispersion for the same cases as in figure \ref{fig:k_width}a. The ratios are computed at very instant of time using the widths shown in that figure, and the corresponding effective mass computed from the instantaneous wavepacket's mean momentum.}
\label{fig:ratio}
\end{figure}

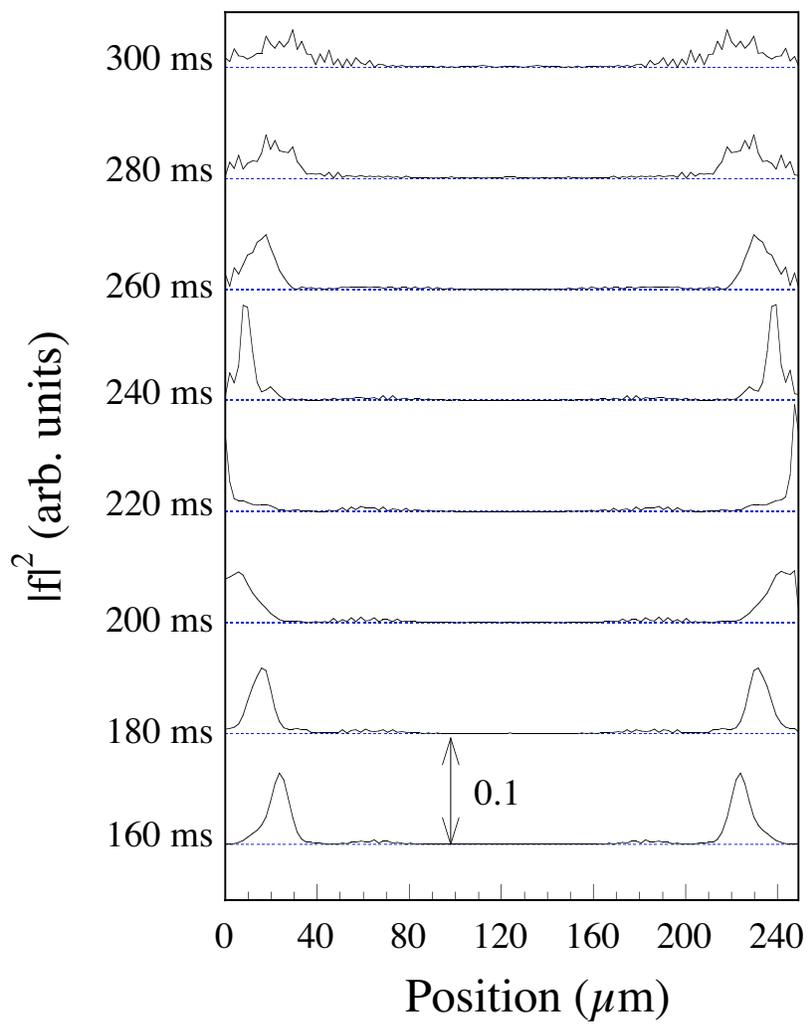
\begin{figure}
\caption{Time evolution of two wavepackets with the same conditions as in Fig. \ref{fig:solit_formation}, but opposite momenta, $-\hbar 58/r_0$ and $\hbar 58/r_0$. }
\label{fig:colision}
\end{figure}


\begin{references}
\bibitem{andre97A} M.R. Andrews {\em et al}, Science {\bf 275} 637 (1997)
\bibitem{ander98A} B. P. Anderson, M. A. Kasevich, Science {\bf 282}  1686 (1998)
\bibitem{morsc01A} O. Morsch, J.H. M\"uller, M. Cristiani, D. Ciampini, E. Arimondo, Phys. Rev. Lett. {\bf 87}, 140402 (2001)
\bibitem{hecke02A} J. Hecker Denschlag {\em et al}  J. Phys. B: At. Mol. Opt. Phys. {\bf 35}, 3095 (2002)
\bibitem{haseg73A} A. Hasegawa, F. Tappert, Appl. Phys. Lett. {\bf 23}, 142 (1973)
\bibitem{zakha71A} V.E. Zakharov, A.B. Shabat, Zh. Eksp. Teor. Fiz. {\bf 61}, 118 (1971) [Sov. Phys. JETP {\bf 34}, 62 (1972)]
\bibitem{khayk02A} L. Khaykovich, {\em et al} Science {\bf 296}, 1290 (2002)
\bibitem{strec02A} K.E. Strecker {\em et al} Nature {\bf 417}, 150 (2002)
\bibitem{burge99A} S. Burger, {\em et al }, Phys. Rev. Lett. {\bf 83}, 5198 (1999)
\bibitem {densc00A} J. Denschlag {\em et al}, Science {\bf 287}, 97 (2000)
\bibitem {carus02A} I. Carusotto, D. Embriaco, G.C. La Rocca, Phys. Rev. A {\bf 65}, 053611 (2002)
\bibitem {alfim02A} G.L. Alfimov, V.V. Konotop, M. Salerno, Europhys. Lett. {\bf 58}, 7 (2002) 
\bibitem {yulin03A} A.V. Yulin, D.V. Skryabin, P. St. J. Russel, Phys. Rev. Lett. {\bf 91 }, 260402-1 (2003)
\bibitem {eierm04A} B. Eiermann, {\em et al}, cond-mat/0402178 (2004)
\bibitem {wai__90A} P.K.A. Wai, H.H. Chen, Y.C. Lee, Phys. Rev. A {\bf 41}, 426 (1990)
\bibitem {jaksc98A} D. Jaksch {\em et al}, Phys. Rev. Lett. {\bf81}, 3108 (1998)
\bibitem {alfim02B} G. L. Alfimov {\em et al}, Phys. Rev. E {\bf 66} 046608 (2002)
\bibitem{gorli01A} A. Gorlitz,  {\it et al.}, Phys. Rev. Lett. {\bf 87} 130402 (2001) 
\bibitem{birk01A} G. Birkl,  {\it et al.}, Opt. Comm. {\bf 191}, 67 (2001)
\bibitem{bongs01A} K. Bongs,  {\it et al.}, Phys. Rev. A {\bf 63}, 031602 (2001)
\bibitem{schmi00A} J. Denschlag, {\it et al.}, Phys. Rev. Lett. {\bf 82}, 2014 (1999)
\bibitem{folma00A} R. Folman,   {\it et al.}, Phys. Rev. Lett. {\bf 84}, 4749 (2000)
\bibitem{plaja02A} L.Plaja, L. Santos, Phys. Rev. A {\bf 65}, 035602 (2002)
\bibitem{ziman} J. M. Ziman, {\em Principles of the Theory of Solids},  Cambridge University Press (1979)
\bibitem{dobre99A} \L. Dobrek {\em et al}, Phys. Rev. A {\bf 60}, R3381 (1999). 
\end{references}
\end{document}